\documentclass{revtex4-2}

\usepackage[outline]{contour}

\usepackage{amssymb,amsfonts,amsmath} 
\usepackage{graphicx}
\usepackage{color}
\usepackage{graphics}
\usepackage{bm}
\usepackage{booktabs}
\usepackage[breaklinks=true]{hyperref}
\hypersetup{bookmarksnumbered, pdfpagemode=UseOutlines, 
pdfdisplaydoctitle, 
colorlinks=true, citecolor=blue, filecolor=blue, linkcolor=blue, urlcolor=blue}

\usepackage[table]{xcolor}
\definecolor{colorp1}{rgb}{0.95,0.95,0.95}
\definecolor{color0}{rgb}{0.8,0.8,0.8}

\definecolor{colormc}{rgb}{0.5,0.2,0.5}

\newcommand*{\affilaachen}{Institut f\"ur Theorie der Statistischen Physik, RWTH Aachen University and JARA-Fundamentals of Future Information Technology, 52056 Aachen, Germany}

\newcommand*{\affilmpsd}{Max Planck Institute for the Structure and Dynamics of Matter, Center for Free-Electron Laser Science (CFEL), Luruper Chaussee 149, 22761 Hamburg, Germany}

\newcommand*{\affilFI}{Center for Computational Quantum Physics, Simons Foundation Flatiron Institute, New York, NY 10010 USA}

\newcommand*{\affilNano}{Nano-Bio Spectroscopy Group and ETSF, Universidad del Pa\'is Vasco UPV/EHU- 20018 San Sebasti\'an, Spain}

\renewcommand{\thesection}{\arabic{section}}

\makeatletter
\renewcommand{\p@subsection}{}
\renewcommand{\p@subsubsection}{}

\makeatother

\begin{document}

\title{
A New Era of Quantum Materials Mastery and Quantum Simulators In and Out of\hfill\phantom{h}\\ Equilibrium\hfill\phantom{h}
}

\author{Dante M.~Kennes}
\email{dante.kennes@rwth-aachen.de}
\affiliation{\affilaachen}
\affiliation{\affilmpsd}
\author{Angel Rubio}
\email{angel.rubio@mpsd.mpg.de}
\affiliation{\affilmpsd}
\affiliation{\affilFI}
\affiliation{\affilNano}

\date{\today{}}

\begin{abstract}

\end{abstract}


\maketitle




{\textbf{Abstract} We provide a perspective on the burgeoning field of controlling quantum materials at will and its potential for quantum simulations in and out equilibrium. After briefly outlining a selection of key recent advances in controlling materials using novel high fluence lasers as well as in innovative approaches for novel quantum materials synthesis (especially in the field of twisted two-dimensional solids), we provide a vision for the future of the field. By  merging state of the art developments we believe it is possible to enter a new era of quantum materials mastery, in which exotic and for the most part evasive collective as well as topological phenomena can be controlled in a versatile manner. This could unlock functionalities of unprecedented capabilities, which in turn can enable many novel quantum technologies in the future.     } 

\section{\NoCaseChange{Preamble}\hfill\phantom{h}}
\label{pre}

Out of equilibrium phenomena are ubiquitous in many fields of science from materials to chemistry to bioscience. They dictate how a system reacts to specific stimuli and how the nonlinear interactions between system and driving can realize effects that cannot be achieved otherwise. In solids entirely novel states of matter may be unveiled under driving conditions. However, those states tend to have finite lifetimes determined by many different parameters. In fact, much like the phenomena of biological life itself, a  nonequilibrium quasi-stationary state in a solid requires constant ``feeding'' by driving to evolve and survive. 

Over the last decades the authors, coworkers and many other experimental and theoretical groups worldwide  have devoted their  research attention towards unraveling the microscopical mechanism behind this fascinating world of driven materials with properties which can be switched on demand by external stimuli. In these materials, many different competing orders and degrees of freedom do play a crucial role necessitating a close consideration of correlations between many particles with different statistics and interaction strengths (electrons, ions, photons) -- and alike all quasiparticle hybrid-states, that can form (such as polaron, polariton, magnon, plasmaron, plexciton, ...) and which can lead to new components out of which to build materials properties in a lego-like fashion \cite{Basov2017,Tokura2017,RubioReview1,RubioReview2,RMPColo,Keimer2017,narang_topology_2021,Tokura2019,CavityMiniReview,Kennes2021}.

This essay will provide a personal and clearly biased view of the authors on how science has progressed through this  scientific journey (we apologize in advance if some contributions are not properly recognized, we want to provide our personal view of the field and its evolution and not to present an exhaustive review of the  whole field). We outline the abundance of new challenges and exciting emergent new physics that came along with the journey so far and that we are just beginning to explore the exciting road ahead. We highlight some of our own humble contributions and what we think the future will bring focusing on -- as the title of this essay suggests-- "mastering materials".
The theater stage of this essay, if you will, is thus one of a quantum theater. In this the main actors for us are electrons and ions (neutrons and protons). The directors here, which advise the actors on their actions, are light fields (photons) or other forms of external stimuli such as pressure or static electric fields. The audience are novel ultrafast time-resolved spectroscopic tools, which probe the changes to the materials' properties. However, when we bring together directors (e.g. photons) and actors (e.g. electrons) in the quantum realm we can realize fascinating matter-light-hybrid states giving rise to emerging, highly tunable polaritonic quantum matter both in their ground and excited quantum state.

This essay starts with a brief introduction and recap of some aspects of the field of research of nonequilibrium material phenomena.  This summary is very contextual and far from exhaustive. It is simply supposed to set the right frame, following the famous quote \begin{center}
    ``{\it If I have seen further it is by standing on the shoulders of giants.}''\;\;\;--Isaac Newton
\end{center}
from a letter to Robert Hooke in 1675, introducing some background on which to rest the discussion.


\section{\NoCaseChange{Introduction}\hfill\phantom{h}}
\label{intro}

Controlling nature has always been one of the pragmatic driving forces of physics and human kind in general. As such, striving for control  proceeds by far  the emergence of modern quantum mechanics and its application in terms of quantum information science, which is raising tremendous attention at the moment for it could revolutionize many technological sectors. This complements the fundamental goals of science which are to foster our basic understanding of nature and elevates them also to a pragmatic level. However, it should not be forgotten, that it is this basic understanding that allows us to exert control over physical systems, explore their novel functionalities and finally in turn realize innovational technologies after all. In this sense it is frequently true that 
\begin{center}
    ``{\it insight must precede application.}''\;\;\;--Max Planck
\end{center}
Along these lines,  obtaining novel forms of  control over {\it quantum many-body systems} and therefore a deepened understanding of these systems is currently a major research objective as it enables rising quantum technologies with unprecedented capabilities. 
First applications of quantum technologies have already  let to revolutionary new developments in the past and are expected to yield even more profound advancements in the future \cite{Preskill2018quantumcomputingin}. Here the {\it objective of control} is central as quantum physics by its very nature tends to have a tendency to escape strong control paradigms. Obtaining a firm grasp on truly quantum properties such as entanglement, collective emergent behavior as well as collective quantum coherence is a -- as Feynman has put it -- ``golly [...] wonderful problem, because it doesn't look so easy'' \cite{Feynman1982}.

In solids quantum control relies in large parts on so-called {\it quantum materials}, which are systems  hosting a wide range of {\it emergent collective and/or topological phenomena}  and which are integral in realizing such next-generation technologies \cite{Keimer2017,Tokura2017,doi:10.1146/annurev-conmatphys-031218-013712} including quantum computing \cite{Smith_2020QM}. The term ``quantum material'' might seem like a misnomer as in reality all materials rely on quantum mechanics. However, the phrase was introduced \cite{Keimer2017} to differentiate materials whose specific properties are rooted on the fundamental laws and quantization of quantum mechanics evading a quasi-classical interpretation from those materials whose properties can still be understood in such a quasi-classical approach. As a consequence quantum materials  have recently moved into the center of modern condensed matter and quantum technologies research. Challenging our theoretical understanding, the fascinating phenomena found in quantum materials, ranging from unconventional superconductivity to topologically protected edge modes, emerge from a delicate interplay between spin, charge, lattice, and orbital degrees of freedom, as well as from the geometric aspects of their wave functions \cite{Keimer2017,narang_topology_2021,Tokura2019}. Recent technological progress on fabricating quantum materials with particular, desirable qualities has advanced the field of {\it materials science} into a new era \cite{Basov2017,Tokura2017,RubioReview1,RubioReview2,RMPColo}, where mastery over the quantum properties is at the vanguard of the current attention. The combined major breakthroughs in  materials science,
 in computational physics 
and in the field of creating light sources 
have further let to giant leaps in understanding materials, their quantum aspects as well as their  interplay with driving forces.  At the interface of materials science and the science of light, novel subfields are born at an exciting rate. This has given rise to the ideas of quantum materials engineering using light, controlling properties by heterostructuring of 2d materials and playing with their stacking arrangement (such as including a twist between one layer to the next) as well as cavity controlled chemistry and cavity materials engineering \cite{Huebener2021,BasovAsenjoGarciaSchuckZhuRubio,PhysRevLett.126.153603,Latinie2105618118,Ruggenthaler2018,Francisco21,Cyriaque21,RubioReview1,RubioReview2,CavityMiniReview,Kennes2021}. Although, still in its infancy these concepts have already brought forth intriguing demonstrations of quantum materials mastery in parts by using external stimuli.

In the same spirit of advancing the quantum-control paradigm, two major trail-blazing research directions have very recently emerged which will be the focus of this essay: (i) (twisted) van der Waals materials (also called moir\'e materials) as novel platforms of quantum materials engineering and (ii) nonequilbrium driving, most prominently by light-matter coupling in vacuum and in and out of equilibrium in a cavity environments-- to access the physics beyond the restrictive linear-response, quasi-equilibrium regime. In particular combining the two, as will be discussed, will unlock light-matter hybrids in two-dimensional twisted heterostructures (with 2 or more layers \cite{liu2020disassembling,doi:10.1126/sciadv.abe8691,Xian2021}).

\subsection{The Rise of (Twisted) van der Waals Heterostructures\hfill\phantom{h}}
\label{intro:vdw}

It is not by accident that early archaeological periods (Stone Age, Copper Age, Bronze Age, Iron Age) are often named after the materials  used in state-of-the-art technologies at the time. It emphasizes the role material discoveries have always played in the ways and fails of civilizational development; in the early times resting mainly on metallurgy. This is a tradition that, although reaching back as far as the beginning of civilization, continues to this day. Modern technological breakthroughs are still enabled by innovational materials platforms and there is no reason to assume the dawning of the quantum information age will be any different. In fact, recently the potential dawning of the topological age, in which  functional topological quantum materials take center stage, has been put forward \cite{doi:10.1063/PT.3.4567}. Our rapidly growing understanding of the fundamental building blocks of matter has led to a novel way of materials engineering: A bottom-up approach in which properties of solids can be engineered at will by growing crystals in a predefined manner. This hope fuels the fields of solid state physics, chemical physics as well as materials science working hand in hand on a deepened understanding of the properties to expect for a given chemical composition of a solid as well as their experimental realization. One line of thought along this direction which emerged are atomically-thin two-dimensional materials. Sparked by the landmark discovery of graphene (awarded the Nobel prize of physics in 2010), two-dimensional materials, which were prior to this discovery believed not to be able to exist \footnote{The Mermin-Wagner theorem \cite{PhysRevLett.17.1133} forbids spontaneous breaking of a continuous symmetry at finite temperature. Therefore, crystalization (breaking the continuous spatial translation symmetry)  is forbidden in two-dimensions and indeed free-standing two-dimensional materials will buckle. However, being supported on a substrate the world of two-dimensional materials becomes accessible.   }, have quickly become a crucial scientific pillar of materials science. Building a three dimensional crystal out of specific two-dimensional layers  have raised the hope that the crystal properties can be controlled in a fully novel way. The building blocks, that are atomically-thin van der Waals materials, are rather versatile, with graphene, boron-nitride, different transition metal dichalcogenides (e.g. WSe$_2$, MoS$_2$, ...), phosphorenes (SnS, GeS,...) or group-III chalcogenides (GaS, InSe,...) being just some marked examples \cite{doi:10.1126/science.aac9439}.  The idea is often compared to playing with legos where the building block of crystals can be engineered from different two-dimensional layers by a controlled stacking sequence \cite{Geim2013}.

\begin{figure}[t]\centering
\includegraphics[width=0.5\columnwidth]{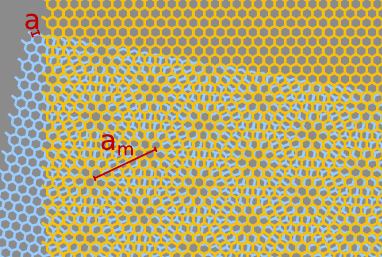}
    \caption{
   {\bf  Moir\'e pattern of two honeycomb lattices stacked at $10^\circ$ angle.}  The lattice constant of one honeycomb lattice $a$ is increased to $a_{\rm m}$. For smaller twist angle the factor of lattice constant increase becomes larger.}
   \label{fig:moire}
\end{figure}

Very recently another impressive arena within the realm of stacked two-dimensional van der Waals materials received tremendous attention \cite{Li2010,Crommie2015,Tutuc2017,Hone2019a,edelberg2019tunable,Cao2018,Cao2018a}. In contrast to lego blocks, which have to be stacked neatly atop each other to hold together and for that purpose feature the same lattice constant on top of each block, van der Waals materials might be staked at a twist \cite{Bistritzer12233,Li2010,Balents2020,Kennes2021,Andrei2021} or different materials might feature different lattice constants, both of which would lead to a (classical geometric) effect known as a {\it  moir\'e pattern}. Dreaded by anyone who works in front of a camera and has a liking for narrow patterned cloths this effect produces a large scale interference pattern which arises when two similar patterns are stacked at a slight twist or when two patterns deviate only slightly in their periodicity (or which arises when the finite resolution of a camera tries to record a very finely structured textile or other material). When the lattice constant mismatch or the twist is small the periodicity of the interference pattern can be large.  This means that the overall lattice constant can be engineered by, e.g, the twist angle between adjacent two-dimensional materials. For small twist angles or lattice constant mismatches the lattice constant of the combined moir\'e material is tuned form the one of the individual chemical bonds $a$ (typically on the Angstrom ($10^{-10}$m) scale) to the one of the moir\'e pattern $a_m$ (typically on the 100nm ($10^{-7}$m) scale) constituting an impressive 2-3 orders of magnitude enhancement. This is shown for the example of a hexagonal lattice (such as found in sheets of graphene) in Fig.~\ref{fig:moire}. Put more generally, this provides an unprecedented control knob on material properties which is usually not available in conventional solids where the lattice constant is given by chemical properties. In fact, moir\'e physics has been omnipresent in surface science for many decades \cite{SurfaceScience,MATSUDA201820180221} (and references therein) but just now took off with the emergence of two-dimensional heterostructures. This is because the tunable kinetic energy scales of the moir\'e crystals allow to tune the interplay of kinetic with interaction, interlayer or spin-orbit energy scales. This can be utilized to control competing correlated orders and topology in a novel way \cite{Kennes2021}. 


\begin{figure}[t]\centering
\includegraphics[width=0.5\columnwidth]{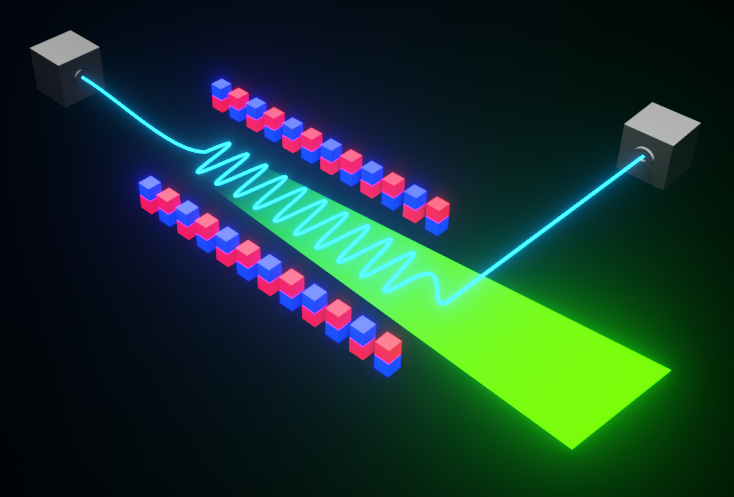}
    \caption{
   {\bf Schematic of a free electron laser.} A particle accelerator is used to create a beam of high kientic energy electrons which are sourced into a magnetic field which will force them on a periodic path. Due to the bremstrahlung  high fluency lasing is achieved.}
   \label{fig:FEL}
\end{figure}

\subsection{Towards Nonequilibrium Quantum Materials Design \hfill\phantom{h}}
\label{intro:Ufast}

In materials science much progress has been made in the last few decades in controlling solids routinely by growth and chemical means (such as introducing dopents, substitutions or vacancies or using different substrates to grow the materials on). Another route is the usage of stacking two-dimensional materials as building blocks of crystals \cite{Geim2013,Dean2010}. While all of these chemical means over the years have led to many fascinating insights and a certain degree of control over interesting material properties, it is still limited in its flexibility, reversibility as well as time dependent controllability. This has recently triggered the quest for control beyond the chemical possibilities of materials  using external stimuli (e.g. light, pressure, strain, electric fields,...), which is a field that has made major experimental leaps in the last few years. On a theoretical level, this requires the description of many-body quantum systems out of equilibrium, which in general poses a formidable challenge. In particular, many materials exhibiting interesting emergent behavior or topological properties show an unprecedentedly strong response to external perturbations, such as pressure or light fields. This on the one hand opens up the route towards efficient control paradigms, but on the other also challenges our fundamental theoretical understanding with many problems in the field being unsolved to this day, triggering a strong research interest in the subject matter. All of this has given rise to the dream of {\it quantum materials on demand} \cite{Basov2017,Tokura2017} which subsumes the broad goal of programming and controlling quantum properties at will by external stimuli, traversing the more conventional idea of chemical control.

One of the triggers of this new research direction has been the development of light sources with unprecedented brilliance, e.g., in free electron lasers building on the technology developed for synchrotrons, which themselves originally played their most prominent role in the completely different field of high energy physics. Although synchrotrons and cyclotrons continue to be celebrated in the field of high energy physics, it has now been realized that the tunable way of creating high energy photonic radiation (light) by accelerated charges has a usage in its own rights: to probe and excite solids beyond the standard paradigm of optics. In a nutshell this new technology relies on synchrotron radiation being generated as an ensemble of electrons is accelerated through a so-called undulator, a magnetic structure sketched in Fig.~\ref{fig:FEL}. For a free electron laser the radiation created is amplified as it bounces through and forth the electron ensemble leading to coherent emission of light; a quantum effect which exponentially increases the brilliance of the laser. With this and other key technologies now at hand, it is time to study the pressing questions of how light and matter interact, how light can control materials on ultrafast time scales (femto- to picoseconds) and what exotic effects may lie beyond the much discussed weak driving regime.
 These advances were matched with impressive developments in the field of ultrafast time resolved electron diffraction to measure changes on very fast time scales \cite{RevModPhys.93.025006,Sie2019}. 

\section{\NoCaseChange{Manipulating Materials with Light}\hfill\phantom{h}}
\label{Mechanisms}

When one aims to {\it probe} the properties of a solid one usually tries to apply a small perturbation and measures how the solid will react. Within linear response, i.e. for perturbations that are very weak, one can relate this response of the solid (described by so-called response functions) to its thermal static or dynamic properties \cite{RevModPhys.74.601,BookLightscattering}. That is to say that the weak perturbation is a probe in the true sense of being able  to generate insights into the equilibrium properties of the solid (without the perturbation present), but does not alter (to any significance) these properties while probing them. 
For decades the linear response paradigm has served condensed matter physicists extremely well and it finds application in almost any interpretation of equilibrium measurements \cite{kubo_statistical-mechanical_1957}. 
However, insightful as this may be, recently, an alternative vantage point has moved into the center of attention, where the probe now is being strongly intensified to such degree that it generates a surgical alteration to the solid 
\cite{Basov2017,Tokura2017,RMPColo,RevModPhys.90.021001,doi:10.1098/rsta.2017.0478,RevModPhys.93.025006,Koppens2014}. In these cases the former probe now acts as a {\it pulse} to the system and properties are being engineered which lie firmly beyond the limitations set by thermal equilibrium (i.e. we are dealing with the non-perturbative response of the system and the previous linear-response paradigm based on perturbation theory in the applied field does not hold). To rephrase, while in thermal equilibrium the control knobs available to scientists are given basically by chemical compositions (synthesis, growth, doping, \dots),  pressure, static electric/magnetic fields, strain and temperature, the nonequilibrium realm offers the tantalizing opportunity of time-dependent control. This upgrades the flexibility of the state of the system from being a thermal one (described by in principle one parameter, temperature) to any state achievable by a nonequilibrium pulse. However, this fascinating paradigm comes with immense experimental and theoretical challenges. 

On the experimental side it is not straightforward to generate such strong perturbing pulses in a controlled fashion.  Recently, advances in light sources received a tremendous experimental push with key developments being made in ultrafast spectroscopies, as well as within creating tailored strong laser sources that can {\it specifically drive desired excitations in materials in an almost surgical way} \cite{Forst2011,Forst2013,GedikARPES1,Mahmood2016,Hu2014,Mankowsky2014a,Kaiser2014,Mitrano2016,Nova2017,kappasalts,K3C60Meta,McIver2020,PhysRevB.102.014311,RMPColo}. As a consequence structural light is one of the leading experimental vehicles to achieve such nonequilibrium control over solids, not least because it comes in many flavors:  We can differentiate between using light in the classical as well as in the deeply quantum limit to achieve control each with and without carrying intrinsic angular momentum; the latter opening up the blossoming field of chiral light sources.  The quantum nature of light  can be relevant in the solid state context when considering tailored environments (e.g in cavities and other environments) connecting to the fields of circuit QED and polaritonic matter \cite{Basov2017,Tokura2017,RubioReview1,RubioReview2,RMPColo,Keimer2017,Tokura2019,CavityMiniReview,Kennes2021,Huebener2021,DemlerWaveguideQED,DemlernonperCav}.

On the flip side, simulation capabilities with respect to nonequilibrium setups have developed significantly now enabling the prediction of novel nonthermal phenomena. 
Such recent theoretical developments allow us to provide guidance to pump-probe experiments, even beyond specific limits such  as slowly or quickly evolving pump fields \cite{Subedi2014,Patel2016,Knap2016,Komnik2016,Babadi2017,Murakami2017,Sentef2017,Bittner2017,Kennes2017,doi:10.1126/sciadv.abd9275,https://doi.org/10.1002/ntls.10010}. Key proposals for control include: (1) exciting the system into a thermally close-by {\it metastable state} with different properties \cite{Stojchevska177,Zhang2016,Teitelbaum2018,Mihailovic19}, (2) the controlled excitation of specific vibrational degrees of freedom, dubbed {\it non-linear phononics}, see e.g. \cite{Forst2011,Subedi2014,Kennes2017}, or (3) the direct coupling of electronic degrees of freedom to a driving light field, called {\it Floquet engineering}, see e.g. \cite{OkaAoki,GedikARPES1,bukov_review,Mahmood2016,KennesFloq,naga_eckstein_floquet_superconductivity_2018,McIver2020,Rudner2020} and will be discussed in more detail below.  The latter is based on the tenet that the dressed quantum states of the material under continuously applied, oscillating perturbation are stationary, endowing the driven material with different properties. The current frontier is to define, classify, and potentially realize strongly interacting Floquet phases and control the strength and form of the interaction.  One of the key recent advances in the field of Floquet engineering concerns cavity engineering to boost the (prototypically small) light-matter coupling by the presence of a mirror cavity. This approach is sometimes refered to as  {\it cavitronics} \cite{Cavity1,Cavity2,Cavity3,Cavity4,Cavity5,Cavity7,Cavity8,Cavity6}. In the latter context, current research efforts address the questions of describing light-matter coupled materials in a cavity from an ab-initio perspective \cite{Ruggenthaler2018}. In the last years a fundamental step towards that goal has been made by merging quantum electrodynamics and density functional theory. This quantum
electrodynamical density functional theory (QEDFT) \cite{Ruggenthaler2018,Flick3026} provides a unique framework to explore, predict and
control light-matter hybrid states which can be created both by driving and by coupling to cavity quantum fluctuations. QEDFT has matured to a valuable tool to address, e.g.  the modification of chemical landscapes for molecules embedded in cavities \cite{PhysRevX.10.041043,Flick2018_moli,Schafer4883}, but describing general materials in cavities coupled to a strong light field still requires better descriptions of the exchange-and-correlation
terms in the theory. In particular, the different intertwined competing energy
scales of light, matter and collective degrees of freedom poses challenges to the numerical description, e.g., because they define time-scales which might vary by orders of magnitude or require the accurate modeling of coupling between lattice and electronic degrees of freedom beyond standard frozen phonon accounts \cite{Pisana2007,Latinie2105618118}. 
Furthermore, the present theoretical ab-initio understanding of quantum cavity dressing is mainly based on the assumption
that the confined photon-field can be described within the dipole approximation, with only very few exceptions even in the simpler realm of model Hamiltonians addressing the issue of relaxing this approximation \cite{DemlernonperCav}.

\begin{figure}[t]\centering
\includegraphics[width=0.5\columnwidth]{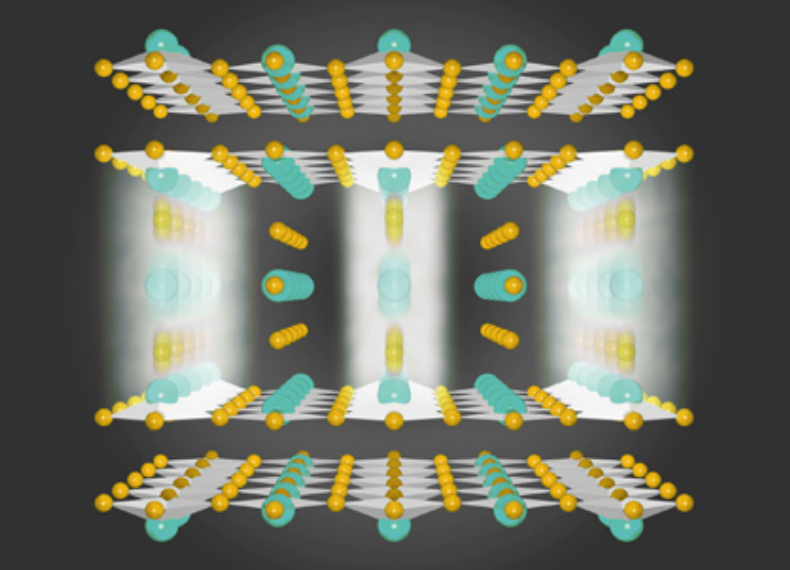}
    \caption{
   {\bf Illustration of Nonlinear Phononics.} Exciting non-linearly coupled phonon modes can be used to change material properties in an ultrafast, dynamic way (figure from  \textcopyright Joerg Harms, MPSD Hamburg). }
   \label{fig:nonph}
\end{figure}
\subsection{Shooting at the Crystal Lattice\hfill\phantom{h}}
\label{Mechanisms:phonons}

 In a series of experiments  \cite{Forst2011,Forst2013,Kaiser2014,Hu2014,Mankowsky2014a}, it was shown that the fragile balance between competing states of matter  can be tuned by exciting the vibrational degrees in a lattice (phonons) using light \cite{Shin2018,Huebener2018_PhononFloq}; see Fig.~\ref{fig:nonph}.  
 In the experiments, an infrared active phonon mode of a crystal is resonantly addressed directly via light in the THz regime. Due to non-linear phonon-phonon coupling between the infrared active mode (coordinate $Q_{\text{IR}}$) to a Raman active mode (coordinate $Q_{\text R}$) of the type 
\begin{equation}
H_{\text{coup}}=\alpha Q_{\text{IR}}^2 Q_{\text R}+\beta Q_{\text{IR}}^2 Q_{\text R}^2,
\label{eq:np_coupl}
\end{equation}   
the lattice is on average stretched along the Raman phonon mode, if we average over the square of the quickly oscillating infrared mode. This mechanism relies on the non-linear coupling of phonons and was consequently dubbed \emph{non-linear phononics}. It was demonstrated that due to this deformation of the lattice the charge density wave order in cuprates was suppressed which in turn unsheaths the competing superconducting order. Theoretical progress has been made in understanding this interplay using density functional theory \cite{Subedi2014} as well as the functional renormalization group \cite{Patel2016}. Ref. \cite{Patel2016} focuses on the theoretical description of competing orders and how their interplay is tuned. It  relies on quasi-equilibrium arguments, which allows for a qualitative account of the experiments, but neglects the dynamic aspect of the problem. In Ref.~\cite{Subedi2014} a static approach within density functional theory is used to map out the free energy landscape  under deformation along the Raman modes ($Q_{\text R}$) from first principles. Along these lines the coupling between the phonon modes Eq.~\eqref{eq:np_coupl} can be determined and treating the modes classically the dynamics resulting from a stimulation of the infrared modes is deduced. A dynamical treatment of a similar system (albeit for a model Hamiltonian), neglecting the quantum nature of the infrared mode, but keeping the quantum nature of the Raman mode can be found in Refs. \cite{Knap2016,Komnik2016,Babadi2017}. Phrased loosely, the authors find amplification of superconducting order by parametrically amplified electron-phonon coupling (between the Raman mode and the electrons). Enhancing superconductivity by light (or any other means for that matter) is of course very intriguing with the ultimate technological goal being room temperature superconductivity. 

The arguably most puzzling and intriguing report of light-induced superconductivity, however, was given in another experiment, where THz light was shone on the organic superconductor K$_3$C$_{60}$ \cite{Mitrano2016,Nova2017}. Here, there is no competing charge density wave order. However, measuring the optical properties in a time-resolved fashion after the pump has been applied the authors report signatures akin to a superconducter at temperatures far above the equilibrium $T_{c}\approx 20 $K on picosecond time scales. At such high temperatures the equilibrium system is metallic.  
Experimentally, superconducting signatures in the optical conductivity persist clearly up to temperatures as high as $100$K suggesting a striking factor of five enhancement of $T_c$ by nonequilibrium control. In further experiments, recently it was shown that light can induce superconducting-like properties in K$_3$C$_{60}$ for much longer times scales \cite{K3C60Meta} and a similar phenomenology was reported in organic salts \cite{kappasalts}  To date the mechanism of this light-induced phenomena is still subject of heated debate: Whether phonons play a role or not or whether the state is indeed a superconducting one (and what the meaning of transient superconductivity really is) remain open (e.g see \cite{kennes_transient_2017,Nava2017,DemlerPerinSC}).

\begin{figure}[t]\centering
\includegraphics[width=0.5\columnwidth]{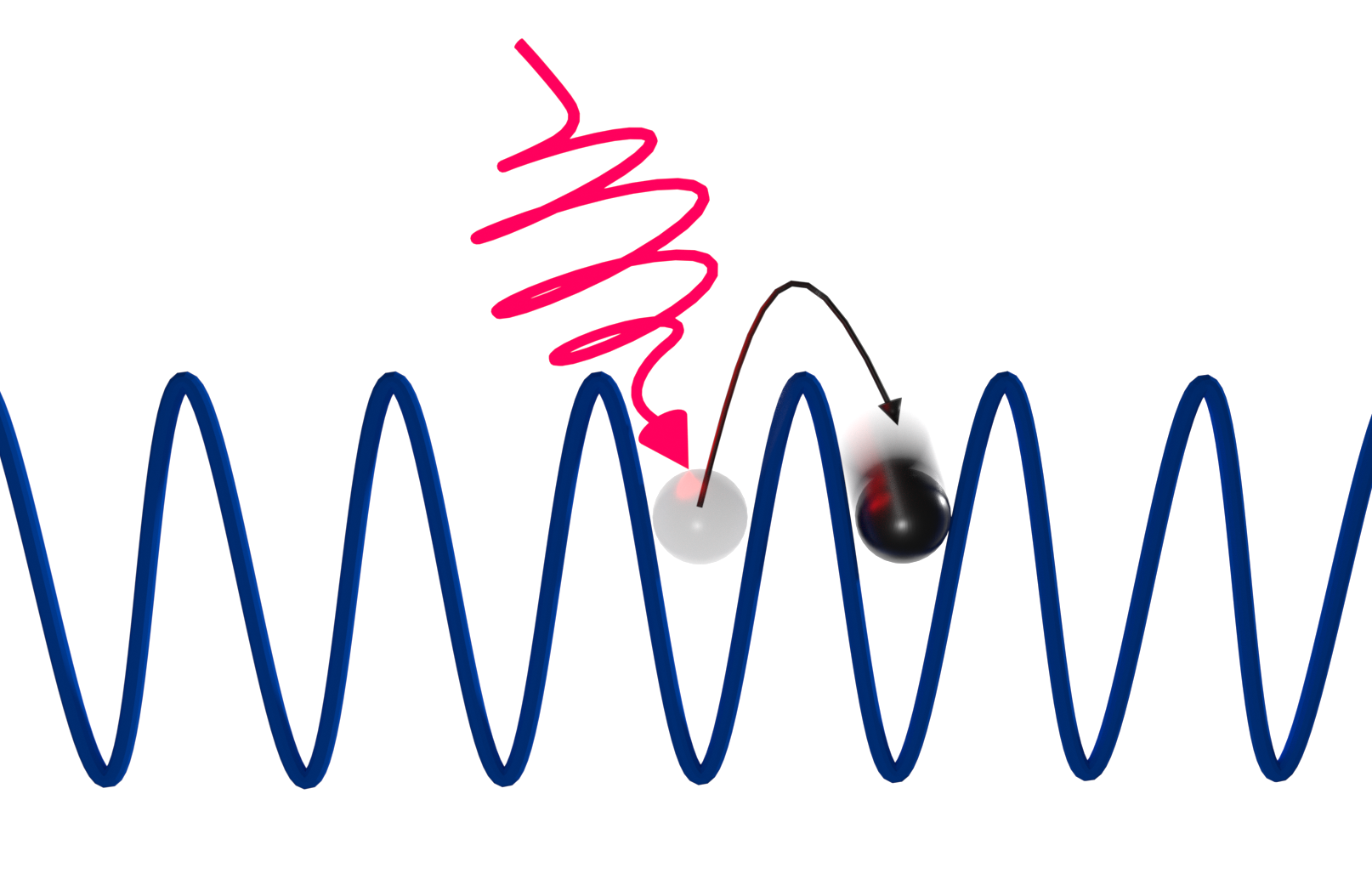}
    \caption{
   {\bf Illustration of Floquet Engineering of Electrons.} Addressing electrons in a lattice (blue periodic potential) by light can be used for Floquet engineering. The electronic Hamiltonian can be altered effectively using the periodic perturbation by the light field.  }
   \label{fig:Floquet}
\end{figure}

\subsection{Shooting at Electrons\hfill\phantom{h}}
\label{Mechanisms:elec}

 On even larger frequency scales then the ones discussed above it is possible to address the electronic degrees of freedom directly. The light field acts as a periodic driving field to the electrons so no additional transduction mechanism like in the above discussed case of non-linear phononics (via lattice vibrations) is needed. This \emph{Floquet engineering}  has emerged as a viral field of physics and embodies studies about how many-body systems can be geared by a periodic drive; see Fig.~\ref{fig:Floquet}. Floquet engineering has already been successfully demonstrated in the field of ultracold gases \cite{Eckardt2017}, but solid state experiments are quickly catching up. E.g., in Refs. \cite{GedikARPES1,Mahmood2016} the authors demonstrate experimentally that Floquet replicas of Dirac cones  can be observed in the band structure using time-resolved APRES studies of the topological insulator Bi$_2$Se$_3$. 
In the pioneering experiment of \cite{McIver2020}, graphene illuminated by a circularly polarized mid-IR pulse was shown to exhibit an anomalous Hall effect.  This is arguably the first experimental solid state demonstration of topological Floquet states \cite{McIver2020}, which contribute to the underlying transport phenomena \cite{OkaAoki,Kitagawa,Sato,Nuske}.  However, although the emergent Floquet–Berry curvature also has a non-zero contribution, one finds that the light-induced Hall effect dominantly originates from the imbalance of photo carrier distribution in momentum space \cite{Sato} in the strong field regime. This finding indicates that intrinsic transport properties of materials can be overwritten by external driving and this may open a way to ultrafast optical-control of transport properties of materials. Another important factor is the  dissipation in the material: Due to dissipation Floquet states cannot be formed in the weak field regime but only once the field strength increases, the photon-dressing effect becomes more significant and overcomes the dissipation effects. In the strong field regime, dissipation acts as to stabilize the Floquet states \cite{Sato}. 
 
 From a theoretical point of view Floquet systems are under heavy investigation (for a recent reviews see \cite{Okareview,Giovannini_2019,RMPColo}). Especially the question of runaway heating in continuously driven interacting systems is of paramount interest. One promising route is to use optimal control theory to put the system into a (close approximant of a) Floquet eigenstate, which does not evolve under drive \cite{AlbertoFloquet}. If runaway heating of a generic interacting system by the external drive cannot be avoided (on the relevant time scales), then in principle the state after long time will be a featureless infinite-temperature-like one \cite{dalessio14,abanin15,mori16,kuwahara16,ho18,abanin17a,abanin17b,mori18,claassen21}. Beyond many-body localized systems \cite{Weidinger2017} at least two regimes where runaway heating can be suppressed were previously identified \cite{bukov15b,Walldorf:AFM,KennesFloq}: (i) the ``Magnus regime" of high frequency as well as (ii) frequencies well within the energy gap of the undriven system (if the system is gapped). In the former, an effective language where the Hamiltonian is replaced by its time average  can be used (or one can include higher orders in this Magnus expansion to go beyond this \cite{Magnus1954}). This allows one to steer some of the electronic properties in interesting ways. Even more flexible control can be obtained in the latter regime, where it was shown that for a Hubbard type system the effective magnetic interaction can be reversed in sign \cite{Mentink2015} or that charge density wave order can dynamically be stabilized or destabilized at will \cite{KennesFloq}. In the same frequency regime, where electrons can be addressed directly, an arguably even more intriguing  way of obtaining quantum control is to directly  address the electronic wave function to tune topological properties (note that the experimental examples \cite{GedikARPES1,Mahmood2016} discussed above were performed on the topological insulator Bi$_2$Se$_3$ already). When the light matter coupling becomes sufficiently large the systems exhibit strongly-correlated electron-photon eigenstates \cite{RubioReview1}, which can be entangled and disentangled by the light matter interplay with intriguing topological applications ahead. 

\begin{figure}[t]\centering
\includegraphics[height=0.26\textwidth]{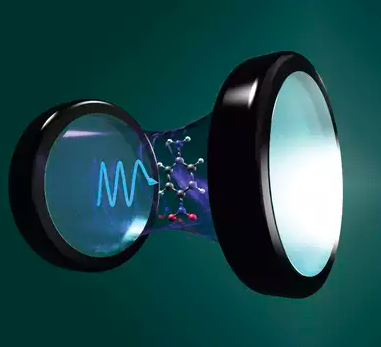}\hfill
\includegraphics[height=0.26\textwidth]{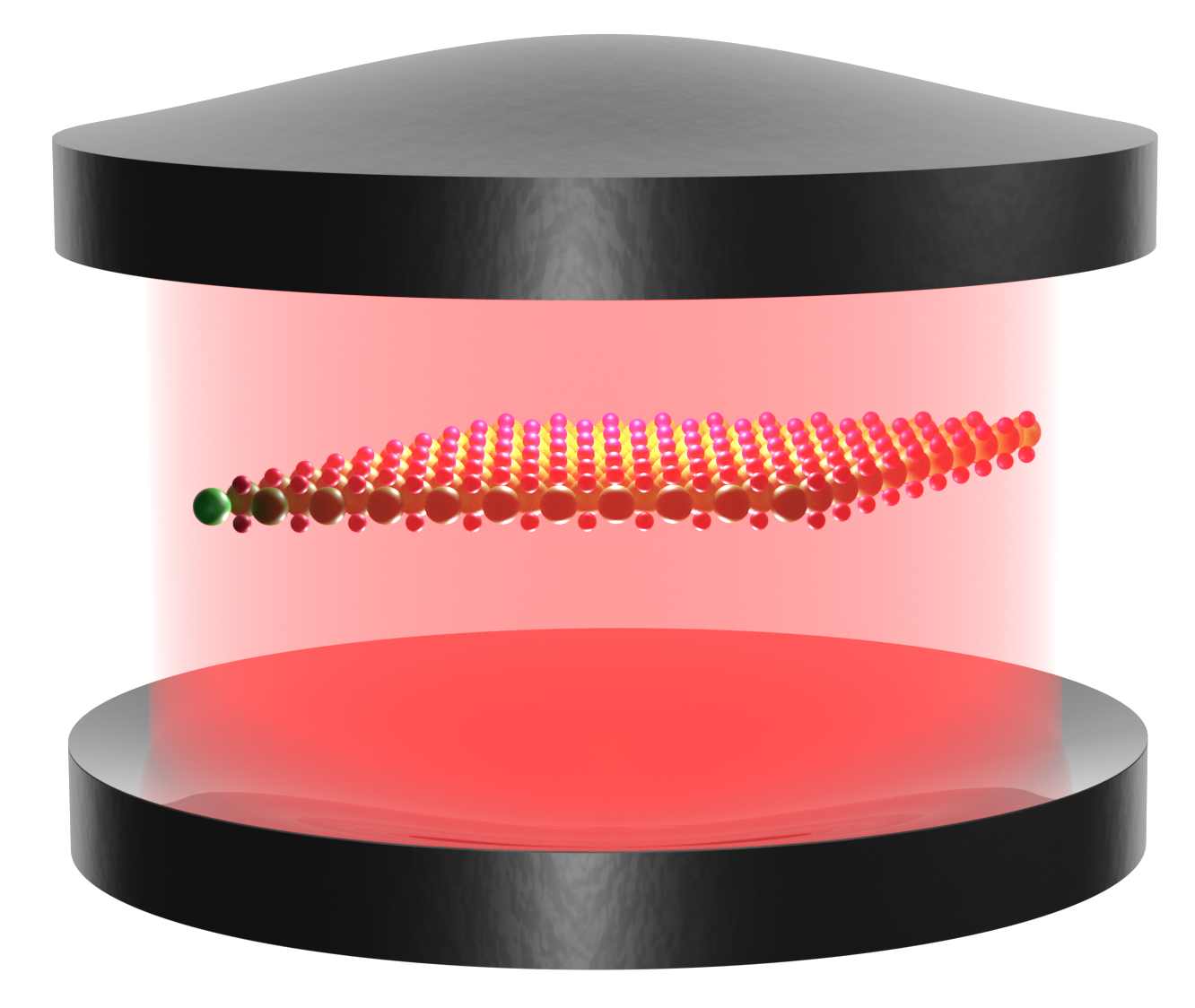}\hfill
\includegraphics[height=0.26\textwidth]{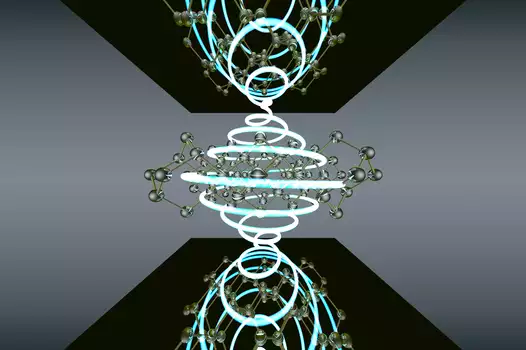}
    \caption{
   {\bf Illustrations of the concepts of Polaronic Chemistry and Cavitronics.} Left: In polaronic chemistry transitions in molecules are effected and investigated by the cavity mode fluctuations. 
   Center: Coupling a material (a two dimensional transition metal dichalcogenide in the illustration) to the vaccum fluctuations of a cavity can alter the material's properties. Right: Engineering the cavity properties, the modes' polarization that couples to molecules or solids can be made circular. This allows to study the influence of chiral light on matter. (left and right figure from \textcopyright Enrico Ronca, Joerg Harms, MPSD Hamburg). }
   \label{fig:cartooncavity}
\end{figure}

\subsection{Amplifying the Light\hfill\phantom{h}}
\label{Mechanisms:cavity}

In the past few decades tremendous efforts have been spend in increasing the brilliance of light sources. One goal is to achieve light strengths for which materials can be controlled using external laser sources. This is a formidable challenge as light-matter coupling, governed by the fine-structure constant $\alpha\approx1/137$, is quite small. Complementary, to this route of increasing the light field strength, very recently an alternative avenue emerged in which one tries to amplify the light-matter coupling itself instead. In the field of cavity-QED \cite{RubioReview2} this secondary route has led to the idea of cavity quantum materials engineering \cite{CavityMiniReview,RubioReview2}, for which material properties are altered by the presence of cavities confining the light field and in turn enhancing the light-matter coupling geometrically; see Fig.~\ref{fig:cartooncavity}. This modification can be obtained either by coupling to the vacuum fluctuations of the light field in such a cavity or by driving the cavity modes with now much weaker laser strengths as the light matter coupling is enhanced.

In this sense, here, we shift the focus on how to realize nonequilibrium states of matter by coupling them instead of to an external drive to the quantized electromagnetic modes in a cavity; the hope here being to realize a novel groundstate (which as such is longlived by construction and no heating problem occurs!) of the coupled cavity-matter system. Nonequilibrium offers the tantalizing opportunity to prepare long-lived metastable states, inaccessible by thermal pathways which can be viewed as (non-global)
 minima in the free energy of the system. Without a drive the system might never reach these states, as the free energy is minimized globally. However, using specific drive protocols, one might prepare one of these states as a long-lived meta-stable state,
 which is hidden from a purely equilibrium protocol. Alternatively, with cavity engineering, the first hope is to stabilize these states with the help of strong coupling to a light field. To give a specific example, it was shown both theoretically \cite{OkaAoki,Kitagawa,Sato} and experimentally \cite{McIver2020} that graphene can experience
 a light-induced anomalous Hall effect upon irradiation by circularly polarized light (which corresponds to the design of the celebrated Haldane Hamiltonian \cite{OkaAoki,Kitagawa}). Dissipation, decoherence and heating of the material play a fundamental role in this nonequilibrium QAH state \cite{Sato}. However, using
 a cavity as a mediator of the QAH effect would remove those unwanted contributions turning the QAH-state into the ground state of the material embedded in a chiral cavity, for example. In other words, this Floquet-to-cavity-QED mapping allows to translate the striking
 properties of the Floquet phases induced by (chiral) light into properties of an equilibrium material in a (chiral) cavity \cite{Huebener2021}.
 Furthermore, by taking the best of both worlds, i.e., to have a (chiral) driving as well as a (chiral) cavity we gain an enormous toolkit to manipulate material properties at will \cite{Huebener2021,PhysRevResearch.2.033033}. The cross-talking that is induced due to the coupled nature of the cavity-matter
 system between the external driving of the different subsystems  has not been explored extensively so far. It promises novel spectroscopic means to generate and control the quantumness of the induced response of the coupled system and to harness the full potential of the
  nature of the light and/or the matter system \cite{Huebener2021}.

Currently the overwhelming majority of theoretical
 studies in the context of polaritonic chemistry and material sciences starts from the dipole and single-mode approximation. The coupling to non-standard light (with angular momentum, breaking of time-reversal symmetry etc.) is then commonly described by adapting
 the dipole approximation in one way or the other. We are  currently pushing the highly ambitious and complementary approach, where we use first-principles QED simulations in full minimal coupling to investigate the effect of cavities on a combined light-matter system.
 Already for the standard linearly-coupled case there are strong differences in observables when comparing the dipole approximation with such a minimal coupling ansatz \cite{Ruggenthaler2018,Flick2019_max,doi:10.1080/00018732.2019.1695875}. 
 While going beyond the dipole approximations for finite systems
 is in principle straightforward, doing so for extended systems leads to several fundamental issues. Full minimal-coupling is not compatible with the periodicity of the Bloch ansatz that is the basis of virtually all solid-state descriptions. Only in the dipole
 approximation in velocity (Coulomb) gauge the periodicity is respected and already for the formally equivalent length gauge picture the usual periodicity of the solid is changed and a novel polaritonic unit cell appears \cite{PhysRevLett.123.047202,PhysRevResearch.4.013012,FreeTB}  leading to a Polaritonic Hofstadter butterfly and cavity control of the quantized  Hall conductance \cite{QEDHofBut}. Recently effects arising from this have raised tremendous attention and led to the demonstration of a modification of the quantized integer Hall conductance in a high-mobility two-dimensional electron gas embedded in a subwavelength split-ring resonator \cite{ModQHECav}. 
 In a nutshell the formation of Landau polaritons in the cavity result in a modifcation of the quantized Hall conductance $\sigma_xy=\frac{e^2}{h}\nu$ with $\nu$ being an integer to  a non-integer $\sigma_{xy}=\frac{e^2}{h}\nu\frac{\omega_c^2}{\omega_c^2+\omega_p^2}$ with $\omega_p=\sqrt{\frac{e^2n_e}{m_e\epsilon_0}}$ and $\omega_c-\frac{eB}{m_e}$ being the diamagnetic shift and cyclotron frequency, respectively. Both the diamagnetic shift and cyclotron frequency depend only on fundamental constants (electron charge $e$, electron mass $m_e$, dielectric constant $\epsilon_0$) as well the magnetic field $B$ and the electron density $n_e$. This observed change in the integer quantum Hall conductivity can only be explained by a cavity-mediated electron hopping via the nonlocal nature of the cavity vacuum fields \cite{ModQHECav}.

Experimental and theoretical research on the control of materials by cavity vacuum fields is destined to considerably accelerate and expand. Controlling the nature of the quantum vacuum surrounding a material inside a resonating cavity can alter its properties imprinting the symmetry of the cavity photons on the matter and design novel quantum materials and phenomena \cite{Basov2017,Huebener2021}. From the theoretical point of view, future possible developments encompass the generalization to spatially non-homogeneous photon modes,  multimode and lossy cavities \cite{CavityMiniReview, RubioReview1,RubioReview2,Ruggenthaler2018,Cyriaque21,Huebener2021}. New predictions for quantum phenomena mediated by cavity photon modes include the enhancement of superconductivity and establishing photon-mediated superconductivity, photon-induced magnetism, ferroelectricity, cavity control of many body interactions in materials and photon-driven topological phenomena in two-dimensional heterostructures and beyond \cite{Kennes2021,RubioReview1,RubioReview2,Bao2022}. Those are among few of the new lines of research that  to be encompassed in the near future under the emerging field of {\it cavity materials engineering}.

\section{\NoCaseChange{Novel Materials and Avenues of Time-Resolved Control}\hfill\phantom{h}}
\label{materials }

As we are developing more intricate quantum technologies the platforms on which to run these technologies also increasingly move into the center of attention. Recently, many breakthroughs in materials science have given rise to the experimental realization of a plethora of different exotic phenomena, many of which build inherently on the principles of quantum mechanics  \cite{QuantumRev2}. In this section we complement the brief excerpt summarizing current light-based control schemes  of the previous section by outlining a selective assortment of topics on such materials-based control, their intrinsic collective behavior and how they interplay with driving.

\subsection{Fabricating New Quantum Materials: A Novel Twist \hfill\phantom{h}}

Stacking layers of van der Waals (vdW) materials has become a flexible avenue towards {electronic band structure engineering} and { marked the rapid rise of vdW heterostructures} \cite{Dean2010,Wang614} in condensed matter physics and materials science. Today, vdW heterostructures are celebrated for their potential in nano-electronics, quantum information and the basic sciences, as the control of band structures allows to tune transport, topological and correlated properties with a high degree of flexibility.   For example, the celebrated Dirac cone at the Fermi level of undoped monolayer graphene grants access to the physics of linearly dispersing electrons. Turning this around, graphene can function as a condensed matter platform to study the physics of ultra-relativistic particles usually described by this kind of dispersion. This builds a bridge between condensed matter and high-energy physics/cosmology allowing to study hallmark questions of high energy physics such as  the Klein paradox \cite{Klein1929}, or by choosing other two-dimensional materials axyon fields \cite{Gooth2019} or Weyl magnetic monopoles \cite{Ma2021}, in a  solid state laboratory. 

To go even further, some of these analogies between high-energy/cosmology and solid state physics even find interesting potential applications such as the Klein paradox enhancing transport properties in graphene \cite{Katsnelson2006}.  
When the stacked layers have a slight lattice constants mismatch \cite{Dean2013}, or are slightly rotated with respect to each other \cite{Neto07,Morell2010,Bistritzer12233,Li2010,Balents2020,Kennes2021}, a long-wavelength periodic modulation is found. This is referred to as a ``moiré superlattice''. A moiré superlattice modifies the electronic bandstructure \cite{Park2008} and can, in selected cases, result in the formation of low-energy sub-bands \cite{Bistritzer12233,Morell2010}. Following such a route, isolated flat bands have been realized in a wide range of graphene-based structures including twisted bilayer graphene (TBG) \cite{Cao2018a, Cao2018, Yankowitz2019,Kerelsky2019, Sharpe605, lu2019superconductors, Serlin2019}, double bilayer graphene \cite{Liu2019, Cao2019, Shen2019,TutucBi,rubioverdu2020universal}, ABC trilayer graphene/BN \cite{Chen2019a, Chen2019, chen2019tunable} and twisted trilayer graphene \cite{Park2021,hao2021engineering}. On the one hand, this { engineering of electronic bands via the moir\'e superlattice can} be utilized to { steer topological properties}  \cite{Chen2020,Polshyn2020,lian2020tbg,stepanov2020competing,Choi2021}. On the other hand, flat band engineering provides a handle to tune the kinetic energy scales to the ones of electronic interactions which can { facilitate control of emergent electronic phases} such as correlated insulators, superconductors and quantum magnets \cite{Cao2018a, Cao2018,Balents2020,Park2021,Kennes2021}. Additionally, the wave functions of the flat bands are topologically tied to the dispersive bands in twisted bilayer graphene which was also found to harbor intriguing effects: For example this topological properties give rise to competing Chern insulating states \cite{stepanov2020competing} as well as unconventional light-matter couplings \cite{prepQuantumgeoTBG}. This flexibility also allows to realize strongly correlated topological phases \cite{Wu2019,chen2020realization}, which are highly relevant for quantum information  \cite{Tokura2017}. 

As briefly stated above correlated phases of matter and topological properties are at the vanguard of condensed matter research for their high potential in key technological applications \cite{Basov2017,Tokura2017}, such as high-temperature superconductivity, high speed and huge capacity memory devices, impeccable security application (in quantum cryptography) or novel information technologies (e.g. in quantum computing). These highly sought after functionalities explains the tremendous interest in twisted vdW materials as candidate systems with highly controllable properties \cite{Kennes2021} (fit for our increasingly information based society  and its need for impeccable security).

\begin{figure}[t]\centering
\includegraphics[width=0.5\columnwidth]{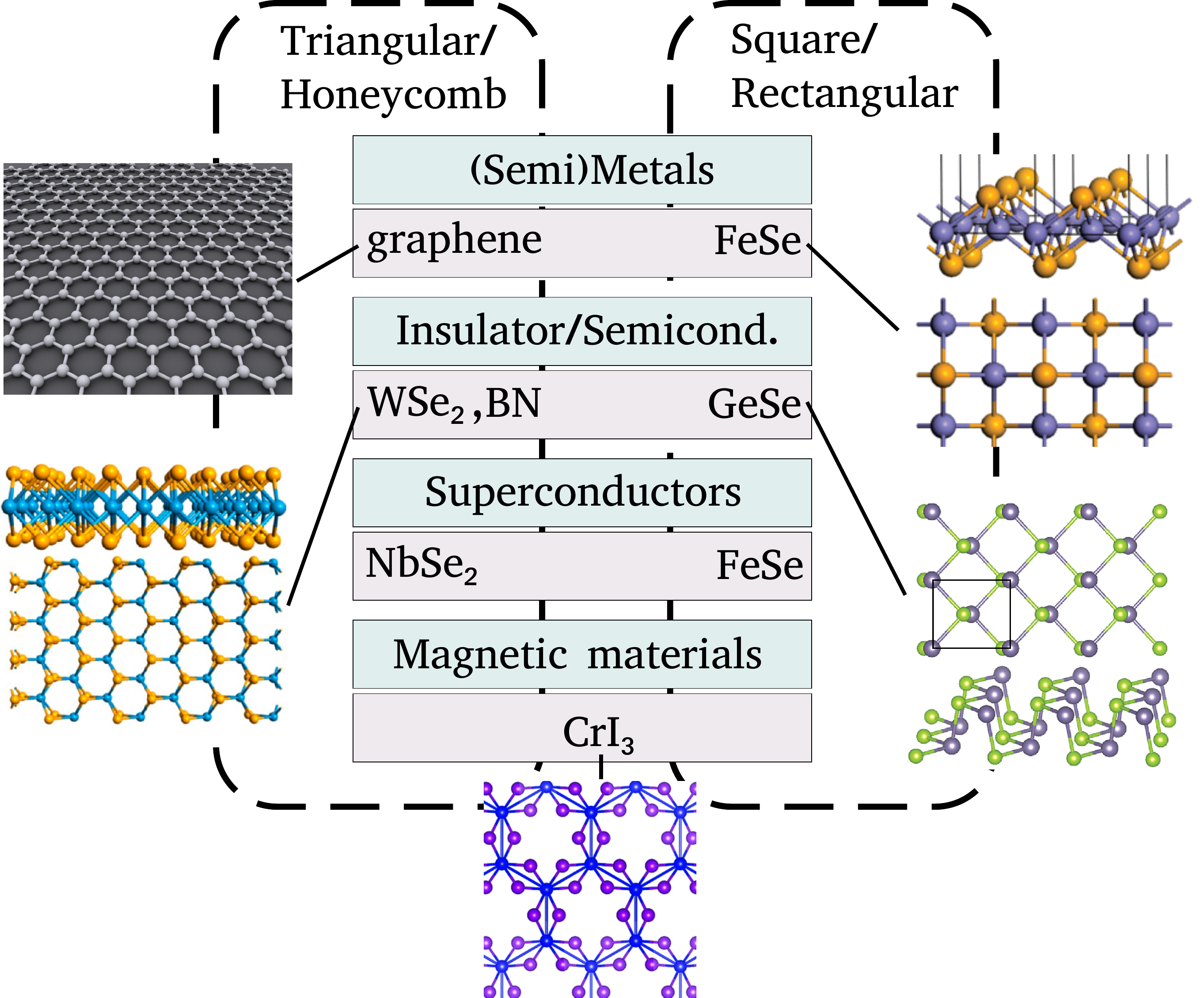}
    \caption{
   {\bf (Incomplete) characterization of twisted vdW heterostructures} via their mono-layer lattice symmetry as well as the phase of matter which they display in equilibrium. Even this very incomplete list hints at the vast combinatoric possibilities offered by twisting these structures.}
   \label{fig:TBG_citations}
\end{figure}

In the most studied case of TBG, the flat bands appear only in certain narrow ranges around specific twist angles, so-called magic angles, the largest of which is $1.1\pm0.1^\circ$ \cite{Bistritzer12233,Cao2018,Yankowitz2019}. This selective appearance arises from  a  delicate interplay between the layer hybridization energy and twist-determined band displacements in momentum space \cite{Bistritzer12233,Cao2018,Yankowitz2019}. The sharp magic angle structure for TBG leads to difficulties in terms of materials design, as slight uncertainties in twist angle result in widely varying band widths and therefore physics. 
Recently, studies of twisted  vdW heterostructures that combine moir\'e patterns with 
using semiconducting  transition metal dichalcogenides (TMDs) \cite{Wu2018, Wu2019, Naik2018, Ruiz-Tijerina2019,Wu2018, Wu2019, Schrade2019,Wang2020WSe2}, twisted bilayer boron nitride \cite{Xian2019BN}  or graphitic systems  where a band gap is induced by electric fields \cite{Liu2019, Cao2019, Shen2019,Chen2019a, Chen2019, chen2019tunable,TutucBi,rubioverdu2020universal}, revealed that in these materials the  bandwidth  varies continuously and smoothly with the twist angle between layers. The absence of sharp magic angles makes these systems less sensitive to small angle variations, which is advantageous for experiments and technological applications in quantum devices.  In  trilayer graphene on boron nitrite  and twisted double bilayer graphene this phenomenon was also demonstrated by using a transverse displacement field to induce a semiconductor bandgap \cite{Liu2019, Cao2019, Shen2019,TutucBi,Chen2019a}. In comparison to these graphetic systems, intrinsic semiconducting  transition metal dichalcogenides (TMDs) provide several potential advantages being less restricted in accessible moir\'e wavelengths and requiring no additional displacement field \cite{Wu2018, Wu2019, Naik2018, Ruiz-Tijerina2019,Wu2018, Wu2019, Schrade2019}.  In this case the bandwidth varies continuously with twist angle, without the need for additional displacement fields, and therefore is widely tunable.

By exploiting the large choice of available TMD materials, properties not present in graphene, such as strong spin-orbit coupling, can be accessed as well. This leads to an additional degree of control to modify electronic properties of twisted materials, and brings  several interesting (correlated) states  of matter within experimental reach, such as Mott states, Wigner crystals, quantum anamolous Hall states among others  \cite{Wu2018,dagotto94,scalapino12,leblanc15,mazurenko17,PhysRevLett.122.246401,Wang2019ex,cao2020nematicity,Christos29543,Li2021,Li2021_b,Li2021_c}. In twisted TMD heterobilayers, such as MoSe$_2$ on WSe$_2$ additionally control over excitons -- bound electron-hole states-- of unprecedented pristine nature as well as their condensation  was demonstrated \cite{Jin2019ex,Wang2019ex,shimazaki2020stronglyex}. 
Going beyond TMDs, alternative materials have been proposed to yield additional control knobs to tune various electronic and structural properties. For example, we have recently shown that twisted bilayers of GeSe can be used to achieve one-dimensional flat bands in a controlled fashion \cite{KennesGeSe}. Alternatively, twisted bilayer MnBi$_2$Te$_4$ was proposed to realize flat Chern-insulators with intriguing topological properties \cite{Lian20}. Even for correlated oxides moir\'e engineering has been demonstrated to alter electronic properties \cite{ChenXinzhong2020}. All of this demonstrates a fascinating opportunity: The catalog of vdW materials is huge with a tremendous potential for control.

This short discussion exemplifies something  profound: While much focus has been put on studying graphitic systems \cite{Tritsaris_2020}, with TBG still at the uncontested center of attention, { there is a huge combinatoric  space of available chemical compositions to exploit which will unlock some of the game changing promises made by quantum materials design}; see Fig.~\ref{fig:TBG_citations}.

\begin{table*}[t]
    \centering
\rowcolors{3}{colorp1}{color0}
\begin{tabular}{ |p{3cm} p{3cm} p{3cm}  p{3cm} p{3cm} p{3cm}|  }
\hline
\multicolumn{2}{|c}{{ \bf Electronic Topology }} &\multicolumn{2}{c}{{ \contour{black}{$\longleftarrow$}\;\;\bf Both\;\;\contour{black}{$\longrightarrow$} }} &\multicolumn{2}{c|}{{ \bf Electronic Correlations }}\\

&&&&& \\

\hline
\multicolumn{2}{|c}{{ Quantum Hall Effect (QHE) \cite{Zhang2005}}} &\multicolumn{2}{c}{{ Fractional QHE \cite{Dean2013}}} &&\\
\multicolumn{2}{|c}{{ Anomalous QHE \cite{Li2021_b,ClaassenZrS2}} } &&&&\\
\multicolumn{2}{|c}{{ Quantum Spin Hall Effect \cite{Wu2019}}} &&&&\\
&&\multicolumn{2}{c}{{ Quantum Spin Liquids \cite{doi:10.1126/science.abi8794}}}&\multicolumn{2}{c|}{{ Quantum Magnetism \cite{chen2019tunable}}} \\
&&\multicolumn{2}{c}{{ Topological Mott Insulators \cite{chen2020realization}}}&\multicolumn{2}{c|}{{ Mott Insulators \cite{Li2021}}} \\
&&&&\multicolumn{2}{c|}{{Wigner Crystals \cite{Smolenski2021,Li2021_c}}} \\
\multicolumn{2}{|c}{{Chern Insulators \cite{stepanov2020competing}}} &\multicolumn{2}{c}{{Correlated Chern Insulators \cite{Nuckolls2020,chen2019tunable}}} && \\
&&\multicolumn{2}{c}{{Fractional Chern Insulators \cite{Xie2021,ClaassenZrS2}}} && \\
&&\multicolumn{2}{c}{{Topo. Superconductors (TSC) \cite{Kennesdwave,PhysRevB.103.024506}}}&\multicolumn{2}{c|}{{ (Un)conventional SC \cite{Cao2018a,Oh2021,Cao2021_triun}}} \\
& &\multicolumn{2}{c}{{Topological CDW \cite{Polshyn2021}}} &\multicolumn{2}{c|}{{Charge Density Waves (CDW) \cite{Zhao2021}}} \\
& &&&\multicolumn{2}{c|}{{Nematic States \cite{rubioverdu2020universal}}} \\
&&\multicolumn{2}{c}{{Topological Luttinger Liquids  }} &\multicolumn{2}{c|}{Luttinger Liquids \cite{KennesGeSe,LLWse2}} \\
\multicolumn{2}{|c}{Higher-Order Topological Insulators \cite{PhysRevLett.123.216803}} &\multicolumn{2}{c}{{Higher-Order TSC \cite{HIgherOTSC}}} && \\
\multicolumn{2}{|c}{Weyl Semimetals \cite{Liu2017}} && && \\
&&\multicolumn{2}{c}{Topo. Excitonic Insulators \cite{Varsano2020}} &\multicolumn{2}{c|}{Excitonic Insulators \cite{shimazaki2020stronglyex}} \\
\hline
\end{tabular}
\caption{{\bf Phases of matter being driven either by topology, correlations or both.} We indicate  a non exhaustive list of (twisted) van der Waals materials related references  where the corresponding phases of matter were discussed. This clearly highlights the versatility of van der Waals materials engineering to realize these exotic condensed matter phenomena.   }
\label{tab:topo_corr}
\end{table*}

\subsection{Exotic Collective Phenomena and their Control \hfill\phantom{h}}

Two thrusts are currently at the vanguard of condensed matter physics: topological and collective phases of matter. 

The former is a relatively young field and relies on the fact that geometric properties of the wave functions can give rise to very robust  novel boundary physics in solids. This robustness --sometimes refereed to as topological protection -- is hailed as an  important ingredient in future quantum technologies.  The arguably earliest discovery of this paradigm is the quantum Hall effect dominated by robust counter-propagating edge modes carrying currents. This effect was later generalized to the anomalous case for which the modes are polarized in one direction as well as to the quantum spin Hall effect for which spin-polarization occurs. In general Chern insulators exhibit the general phenomenology of robust edge modes linked to non-trivial geometric aspects of the wave function. Following this line of thought it is nowadays understood that at an interface to a topologically trivial system (such as the vacuum) a closing of the gap in Chern insulators and a corresponding edge mode is imperative. Soon after the concept of such topological gaps protecting edge modes in topological insulators and superconductors was introduced, multiple classes of systems were identified hosting gapless bulk modes but yet topological edge states are found. One of these classes of systems are Weyl semimetals \cite{Armitage2018,doi:10.1146/annurev-conmatphys-033117-054129,doi:10.1146/annurev-conmatphys-031016-025458}, which host an even number of gap closing points with opposite chirality of the wave functions at these closing points, which are connected by topological edge modes at interfaces to a vacuum. Thus, such a system exhibits features ``in between" of those of a topological insulator (edge modes) and a metallic system (ungapped). The gap closing points, the so-called Weyl nodes, which determine the edge mode properties were demonstrated to be highly susceptible to  chiral light fields \cite{PhysRevB.93.155107,Hubener2017,Bucciantini2017,Deb2017}. To generalize these concepts further, higher-order topological insulators were introduced \cite{PhysRevLett.123.216803,HIgherOTSC}. In higher-order topological system the edge modes are not $D-1$ but $D-n$ with $n>1$ dimensional, where $D$ is the dimension of the system. E.g., here robust {\it corner}  (zero dimensional) states at the surface of two-dimensional systems in contrast to the (one-dimensional) edge states discussed above are found. All combinations of dimensionality of the system and the surface states can be discussed and give rise to a rich playground to control and modify surface properties by engineering the bulk material.  Nowadays, many of these topological phenomena are understood in terms of topological invarians, the most clear example of which is the integer quantum hall effect being characterized by the integer-valued Chern number.

In contrast,  collective phases of matter refers to the physics emerging in systems where particles are (strongly) correlated among each other. This emergence can give rise to entirely novel physics and is the foundation of phenomena such as quantum magnetism and superconductivity, which find widespread applications already. However, the catalog of emergent phenomena extends far beyond these two well-studied cases. They, e.g., include more exotic types of insulators, such as Mott insulators, for which the insulating behavior is driven by many-body interactions, Wigner crystals, for which interaction lead to a spontaneous breaking of the spatial symmetry (crystallization) or charge density wave formations which freeze charge movement in place. It can also give rise to fundamentally important concepts such as Luttinger Liquids in quantum wires, which exhibit power-law dominated transport properties firmly beyond an effective single particle picture or the prediction of excitonic condensates, which are analog to superconductors, but with the paired electrons (cooper pairs) being replaced by condensed particle-hole pairs instead. For the study of these highly complex and intuition-defying phenomena new simulation techniques are brought to life. Complementing the traditional route of characterizing these phenomena in and out of equilibrium using techniques from quantum many-body theory, nowadays there is a strong push towards quantum simulations \cite{Trabesinger2012,bloch2012quantum,Kennes2021}. In quantum simulations, in contrast to the traditional computing approach, a highly controlled quantum system  is realized physically which ought to mimic the original Hamiltonian  under scrutiny. This approach again (see above) relates back to key ideas of R. Feynman \cite{Feynman1982}:
\begin{center}
    ``{\it Nature isn't classical, dammit, and if you want to make a simulation of nature, you'd better make it quantum mechanical.}''\;\;\;--Richard Feynman
\end{center}
In a sense this provides a full loop in our approach as we want to understand the quantum nature of materials, we try to abstract the relevant parts in a model Hamiltonian, which we then again solve using a quantum mechanical simulator \cite{Trabesinger2012}. Intriguingly, as mastery over solids increases the basis of such in and out of equilibrium quantum simulators might move from, e.g., ultracold gases of atoms confined in optical lattices \cite{bloch2012quantum} towards a material platform in certain cases, as we have recently proposed in Ref.~\cite{Kennes2021} in the context of ``a quantum moir\'e simulators''.

At the interface of topology and electron correlations -- as usual in science -- an even more rich catalog of physics is discovered. Strong correlations can give rise to fractionalization of charges, e.g. in the fractional quantum Hall effect or fractional Chern insulators. These systems could harbor intriguing anyonic quasiparticles with braiding statistics going beyond the simpler fermionic ($-1$) or bosonic ($+1$) cases. These particles are believed to be important in quantum information sciences as they could allow topologically protected and universal quantum gate operations. In the context of quantum magnetism, spin liquid phases, phases which avert magnetic ordering and are often characterized by intriguing topological properties are also raising tremendous attention at the moment \cite{doi:10.1126/science.abi8794} just as well as topological superconductors which might host evasive Majorana edge modes. A short summary of the intriguing behavior in topological and/or correlated systems is presented in Table~\ref{tab:topo_corr}.

With this exciting plethora of phenomena at hand (of which we have barely scratched the surface!), which have only partially been discovered experimentally let alone exploited in technologies, it comes to no surprise that scientists are pushing hard on the boundaries of what can be engineered in artificial systems.  With every new material or other avenue of control such as stacking or driving in our tool belt, we come closer to the goal of quantum materials mastery and therefore the realization of the fascinating physics hinted at above.   This plausibilizes the exponential growth in the field of materials science that we witnessed in the last decades. It also explains why twistronics and ultrafast materials science is on the rise. In the end we want to obtain materials mastery on the level of the quantum degrees of freedom and be able to modify topological and collective phases at will, which in the best case will enable new technologies. Twistronics is still very much in its infancy, but initial experiments show the difficulties in obtaining large, clean areas that are unstrained and have a  homogenous twist for which the flat band physics discussed above dominates. In ultrafast materials science the large driving fields required to modify materials properties substantially, come with heating; a major challenge to be overcome in the future. Therefore, new ideas and concepts are required to pave the road ahead.

\section{\NoCaseChange{The Road Ahead: The Future of 2D Materials Is Bright}\hfill\phantom{h}}

The future of two-dimensional materials is bright. As hinted at by the references in Table~\ref{tab:topo_corr} many of the fascinating topological and many-body phenomena prototypical to condensed matter research have already been proposed and/or realized in such material platforms. Yet, there is more! On the one hand,  there are  many additional known phases of matter and phenomena going beyond the short exposition of Table~\ref{tab:topo_corr}  in particular and this brief overview in general which we expect to be  realized in two-dimensional materials in the future. On the other hand, the journey of exploring the consequences of topology, correlations and other unifying themes in condensed matter research are far from complete. Exciting surprises are expected to be unveiled and discoveries to be made in the future, not least building on the extremely versatile platform that is two-dimensional materials. 

We believe that there is also another meaning in which the future of two-dimensional materials is bright: the very promising route of interacting with heterostructures using light in the different versions that we have described in the previous pages. This provides a unique platform in which to explore many interdisciplinary concepts from fields ranging from quantum thermodynamics, materials science, quantum optics, particle physics, cavity-QED, dissipative (non-hermitian quantum) systems, fluid dynamics and more. Many new phenomena at the interface of those disciplines could be realized using this condensed matter platform and would enable a better understanding of the physics (as it becomes relatively easily measurable!). It would help in the development of our understanding of new states of matter that build on those cross-disciplinary concept,  as for example realizations of novel quasiparticle condenses and polaritonic quantum matter (based on light-matter hybrids) become available. In the previous sections we have presented an overview of the impressive achievements in the field of control of nonequilibrium phenomena in materials and the realization of new phenomena and functionalities. We have also highlighted the immense possibilities offered by novel 2D heterostructure platform combined with optical cavities (quantum light) or interacting with a laser field (pump probe studies). We have focused here mainly on the impact on materials but the present findings can have applications to chemistry (control of chemical reactions and energy transfer phenomena)  as well as biophysics.

\subsection{Polaritonic Chemistry\hfill\phantom{h}}
Seminal experimental results have uncovered that by engineering the vacuum of the electromagnetic field via, e.g. optical cavities, chemical reactions can be changed. Since this can happen at standard ambient conditions in solvation, such a novel control technique is very promising for future quantum technological applications. This field of polaritonic chemistry has become a rapidly developing one over the last years. For example, the hybridisation of light-matter
(quantum) states within a cavity, i.e. the formation of polaritons, has lead to astonishing experimental discoveries. It was observed that vibrational strong coupling can inhibit \cite{https://doi.org/10.1002/anie.201605504}, steer \cite{doi:10.1126/science.aau7742} and
even catalyze \cite{hiura_shalabney_2021} a chemical process. Moreover, seminal measurements were published on the control of
photo-chemical reactions \cite{doi:10.1126/sciadv.aas9552}, energy transfer \cite{Coles2014}, the realization of single molecular strong coupling \cite{PhysRevX.7.021014} and even
evidence for the increase of the critical temperature in superconductors was reported \cite{EbbesenSC}.
polaritonic chemistry is a notoriously hard problem to solve from a fundamental theoretical perspective \cite{Sidler2020}. This originates from
the strong hybridization of light and matter, which a priori requires to treat not only the
matter part with "chemical" accuracy, but also the involved  electro-magnetic fields.
The theoretical understanding of such cavity-mediated chemical reactions is very challenging. Besides the well-known complexity of large chemical systems we have also to consider the strong coupling to cavity modes and collective effects. Very recently, first ab-initio simulation methods emerged, which solve the Pauli-Fierz Hamiltonian numerically. Computational
implementations involve quantum electrodynamical density functional theory (QEDFT) \cite{PhysRevA.90.012508} and coupled
cluster theory for molecular polaritons \cite{PhysRevX.10.041043}. Early applications of these novel ab-initio methods indeed suggest that the widespread phenomenological models cannot capture the entire story
of polaritonic chemistry and important effects are absent. Linear response QEDFT has
demonstrated that strong local modifications of the electronic structure emerge in the vicinity of impurities, embedded within a collectively coupled environment. By construction, collective phenomenological
models did not include this effect. The observation of this complex interplay has started a paradigmatic
shift in the understanding of polaritonic chemistry, since it re-introduces the principle of locality for
polaritons \cite{Sidler2020}, which is key for the understanding of chemical processes (e.g. reactions). Furthermore,
QEDFT Ehrenfest dynamics reveal that the cavity can correlate multiple vibrational degrees of freedom, which redistributes energy from a specific bond to other degrees of freedom
and eventually suppresses the bond breaking \cite{Schaefer21}. Still there is plenty of experimental and theoretical work to be done to unravel all miscrocopical and collective processes behind the  control of chemical phenomena in cavities as well as their reliable prediction.

\subsection{High Energy Physics (and Beyond) in a Condensed Matter Lab\hfill\phantom{h}}

The interdisciplinary character of the research discussed so far is also evident when connecting  condensed matter to high energy physics. To provide further examples of  highly intriguing directions of future research (beyond the above), we name  type-I and -II Dirac Fermions in condensed matter systems as a direct link between the low-energy excitation of a Dirac-material in solid-state physics and that of a quantum theory in flat (type-I Dirac fermions) or curved (type-II Dirac fermions) spaces from general relativity.   Another example briefly touched upon in the main text is provided by the chiral anomaly in astrophysics and cosmology. As mentioned analyzing the behavior of  Weyl nodes in materials allows for a solid state inroad into this phenomena.  In certain solid state systems and conditions, it is possible to tune the effective fermion mass experimentally \cite{Cha2010}. Thus, one can study the effectiveness of the chiral magnetic effect as a function of the effective fermion mass. With suitable rescaling of the relevant quantities, one could transfer such measurements in condensed matter to astrophysical environments. Therefore, this route might allow us to address open questions about how effective the chiral magnetic effect can be in astrophysical  environments from a very controlled condensed matter viewpoint.

Strongly correlated phenomena and the emergence of hydrodynamic flow in materials (exotic metals) is an alternative, interesting avenue of research that connects to such a multidisciplinary approach. The tools and advances briefly highlighted in the present essay would enable the control and realization of these states in a real material.

\begin{figure}[t]\centering
\includegraphics[width=0.5\columnwidth]{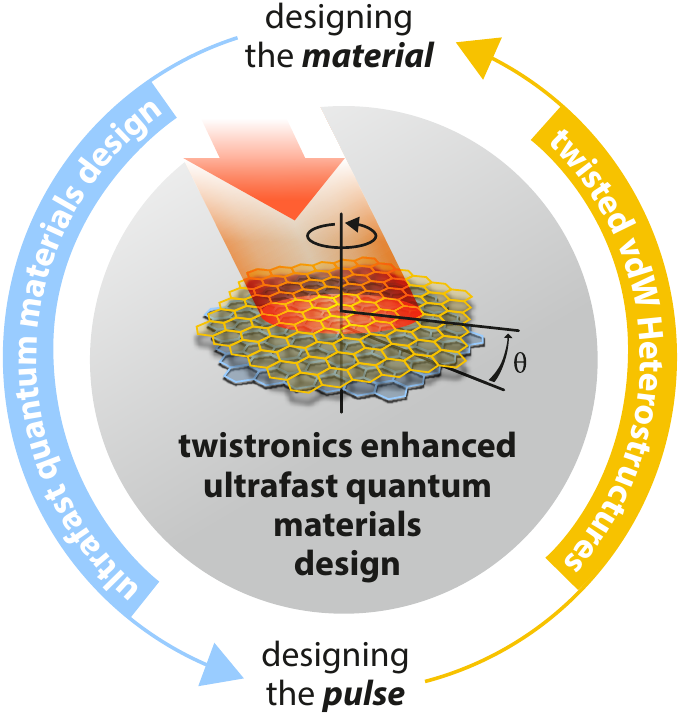}
    \caption{ {\bf Bidirectional optimization approach for driven materials' phenomena} enabled by innovative quantum materials design such as twistronics. In a synergistic fashion the optimization of twisted vdW materials' properties to the drive and vice versa will  advance nonequilibrium control of solids beyond the current limitations set by detrimental heating effects and the limited flexibility of conventional solid state based platforms. }
   \label{fig:bidirec}
\end{figure}

\begin{figure}[t]\centering
\includegraphics[width=0.9\textwidth]{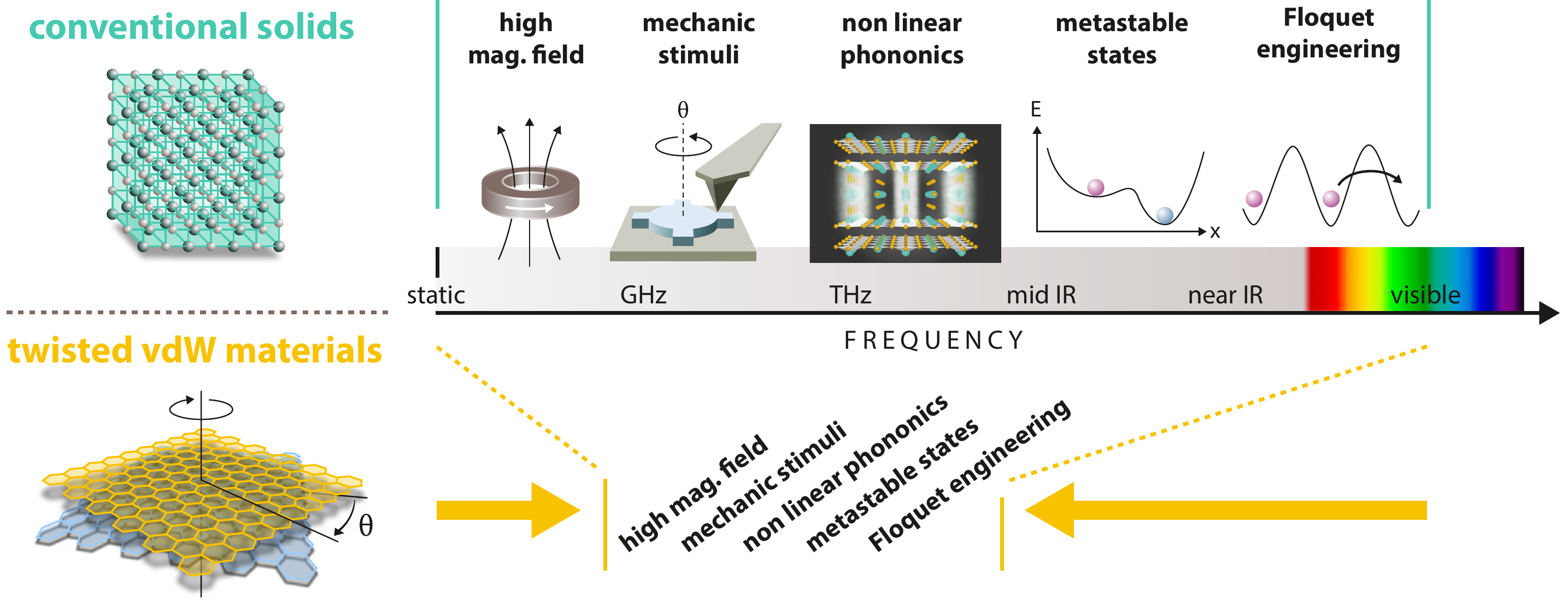}\vspace{1cm}
\hfill\includegraphics[width=0.7\textwidth]{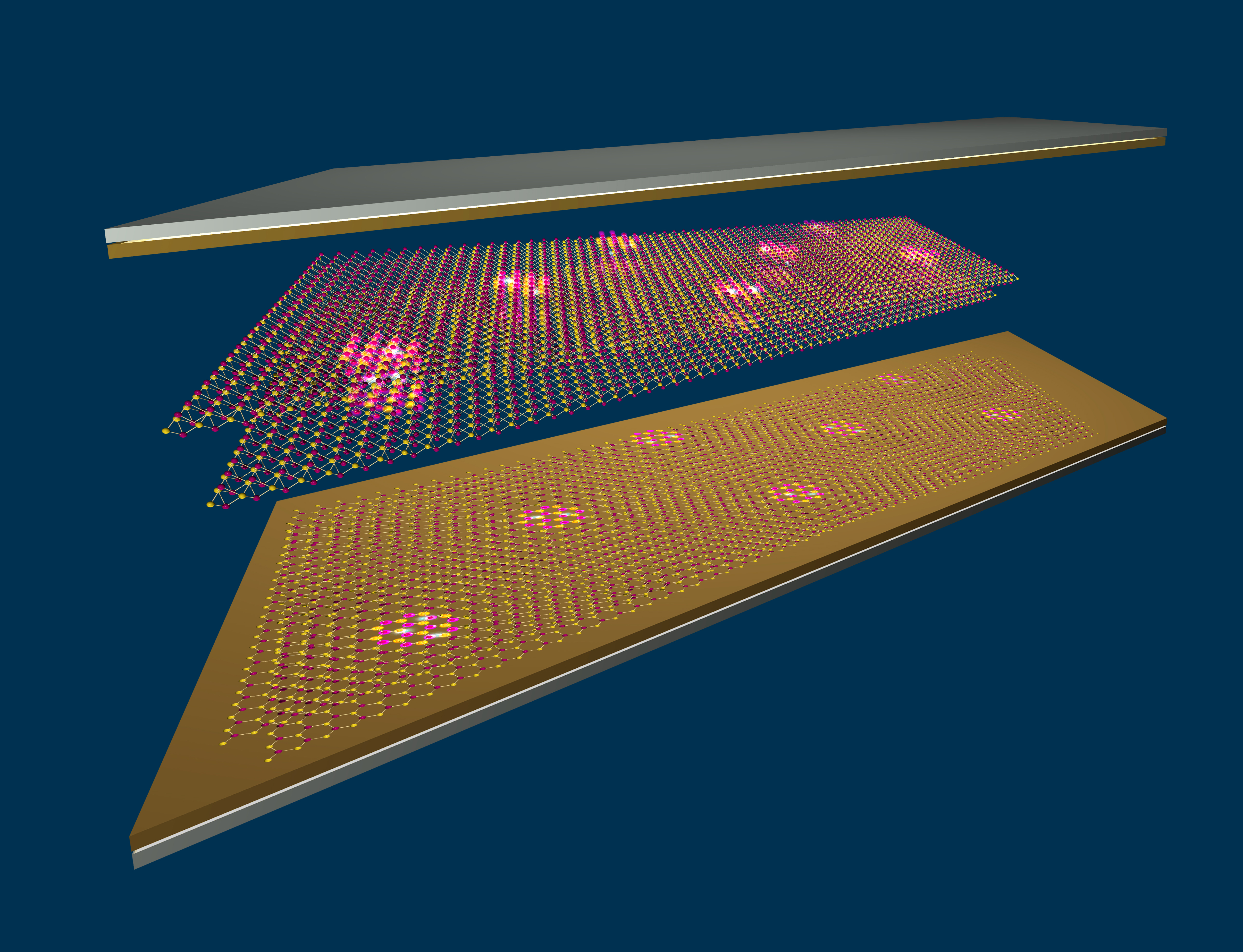}
    \caption{
   {\bf Mechanisms of control in novel quantum materials.} Top: Frequency scale illustrating  different nonequilibrium control routes in conventional solids (top and images in the center) as well as in the case of twisted vdW materials (bottom). For the latter, control mechanisms are squeezed to similar energy scales. Bottom: Illustration of cavitwistronics. By utilizing the advantages of both cavitronics and twistronics novel physical regimes and control forms should be accessible (lower panel from \textcopyright Joerg Harms, MPSD Hamburg).}
   \label{fig:plotmod}
\end{figure}

\subsection{(Quantum) Floquet Materials Engineering\hfill\phantom{h}}

In Floquet materials engineering, beyond a variety of Floquet topological phases, such as Floquet topological insulators and Floquet Weyl semimetals for example, which replicate existing equilibrium analogues, there is a treasure trove of possible Floquet states, which have no equilibrium counterpart. The driven Floquet state, however, is hard to realize experimentally in materials because processes such as dissipation, decoherence, impurity scattering, excited state lifetimes and heating strongly limit the survival of the Floquet states. Furthermore, the actual population of Floquet states is specific to the driving protocol. As we discussed in the subsection \ref{Mechanisms:elec}, 
 especially the question of runaway heating in continuously driven interacting systems is of fundamental interest as it, if it cannot be avoided, dictates a featureless infinite-temperature-like state at long-times. One direction to avoid this, which we believe will be of major relevance in the future is the Floquet-to-cavity-QED mapping \cite{Huebener2021}. This generalizes the concepts of polaritonic chemistry towards the solid state realm. The major theme here being that controlling the nature of the quantum vacuum acting on a material inside a resonating cavity allows to steer some of the electronic properties in interesting ways. Despite the considerable theoretical difficulties in describing the driven-dissipative nature of these hybrid light-matter correlated systems, the developments have fueled theoretical work to not only explain the experimental observations, but to lay out a set of proposals and to define an agenda for strongly-correlated electron-phonon-photon systems (which can led to new light-matter hybrids in interacting polaritons).

Furthermore, cavity-modes can both mediate materials’ correlations and be an additional knob to measure and control electronic systems that are interacting from the outset. Such photon mediated interactions can be engineered to be drastically different and induce long-ranged interactions and non-local entanglement. When matter with intrinsic strong electron correlations is embedded in a cavity, qualitatively new phenomena are expected to emerge as a whole zoo of polaritonic quasiparticles come to life. Just like electron-hole pairs (excitons) can be hybridized with light, various collective modes (e.g. Goldstone modes, Higgs particles, Bogoliubov quasiparticles, etc.) can be mixed with photons leading to non-local composite polaritons \cite{BasovAsenjoGarciaSchuckZhuRubio}. Driving the cavity can induce condensation of this new form of light-matter hybrids with strong and tuneable interactions, paving the way to quantum control of coherent phases.

\subsection{Twistronics for Ultrafast Quantum Materials Design\hfill\phantom{h}}

In the previous sections, we have introduced the concept of moir\'e van der Waals heterostructures as robust solid-state based quantum simulation platform to study strongly correlated physics and topology such as Mott and fractional Chern insulator, ferromagnetic and correlated QAH insulator, quantum chiral spin liquid, superconductivity (d-wave, triplet pairing, in a Kitaev lattices,...), Majorana fermions, and many more \cite{Kennes2021}. A high-impact direction of future research is to use this novel materials platform to enhance the capabilities of ultrafast quantum materials design. We believe this to be a particularly fruitful direction as the electronic band structures of moir\'e crystals can be engineered in a particularly flexible manner. As mentioned, currently one of the main limitations of the field are  heating effects, which are detrimental to control. This is where  moir\'e engineering of materials can intervene, holding the timely promise that the properties of these twisted vdW materials can be controlled with high flexibility, allowing for a unique  bidirectional optimization, see Fig.~\ref{fig:bidirec}, of the driving and of the material towards each other (the current state-of-the-art is mainly to engineer the driving towards the material due to lack of material control in conventional solids). Furthermore, in moir\'e materials different control mechanisms are moved to comparable energy scales utilizing the  twist-induced reduction of the kinetic energy; see top panel of Fig.~\ref{fig:plotmod}. This allows us to address and exploit them in a synergistic fashion.  The methodologically  ambitious goal of   combining the fields of twisted van der Waals materials with the one of ultrafast quantum materials design is thus particularly timely and synergistic in nature. It advances nonequilibrium quantum materials design into a new era, beyond the current limits set by the lack of device flexibility and by  heating effects.

Finally, we propose ``Cavity Twistronics'', i.e. to combine cavity quantum electrodynamics (QED), quantum reservoir engineering and 2D-twisted van der Waals heterostructures to build a novel and unique platform that enables the seamless realization and control of a plethora of interacting quantum phenomena, including exotic, elusive and not-yet envision correlated and topological phases of matter. An illustration is shown in the bottom panel of Fig.~\ref{fig:plotmod}. This research direction will open up exciting new opportunities for the study of the interaction of quantized light with quantum many body systems with a number of intriguing observations and few exotic predictions ranging from enhancing superconductivity \cite{Cavity4,EbbesenSC} to driving paraelectric-ferroelectric transition in perovskites \cite{Huebener2021,PhysRevX.10.041027} to control of excitons in 2D materials \cite{latini18} to strongly correlated Bose–Fermi mixtures of degenerate electrons and dipolar excitons \cite{paravicinibagliani18} to emergent topological phases. Cavity control of many-body interactions in strength and form, realized in these highly tunable platforms is an particularly interesting direction of Cavity-QED. Complementary, cavity enhancement of electron-phonon coupling \cite{Cavity4} could strengthen interactions and provide unique means to control properties of moir\'e bands. This facilitates the microscopic understanding of superconductivity and other strongly correlated phenomena inside a cavity.
    Moving towards 3D material’s embedding we can envision stackings of 2D materials to be a promising route. E.g. stacks at alternating twist meaning that the twist angle alternates between two values, zero and $\alpha$, could be particularly promising \cite{Xian2021}. If $\alpha$ is small, in-plane localized sites will emerge by the moiré interference, and these sites will lie directly atop each other in the out-of-plane direction We expect this to give rise to a 3D quantum Hall state. Utilizing the cavity modes as synthetic dimensions we could realize a 4D-quantum Hall effect \cite{doi:10.1126/science.294.5543.823} and even go to effectively higher dimensions. 
    
    Looping back to the connection to other fields of physics, highly engineered moir\'e materials hold further promises to extend concepts of high-energy physics. For example, entirely
      new fermionic quasiparticles beyond those known from high-energy physics could be engineered in cavities.  In quantum field theory, three types of fermions play a fundamental role in our understanding of nature: Majorana, Dirac, and Weyl. They are constrained by the Poincare symmetry in high-energy physics. In contrast, electrons and quasi-particles in materials obey the space group symmetries, which are less constrained. Therefore, beyond  Dirac, Weyl and Majorana fermions, new fermions with non-trivial topological properties could be observed \cite{doi:10.1126/science.aaf5037}. Inspired by this idea, and using the cavity modes as additional synthetic dimensions, we propose to extend the theoretical group symmetry classification to the larger light-matter hybrid space to identify conditions under which spin-1 and spin-3/2 fermions as well as other (fractal) fermions can be created. 

\section{\NoCaseChange{Closing}\hfill\phantom{h}}
In closing, the field of manipulating quantum materials with light is, in our humble opinion, destined to yield many exciting surprises in the coming years. We hope that this has become apparent from the collection of topic we chose to discuss above. However, the reader should be advised that the list of phenomena and routes of future research we provided is not complete. Quite the opposite is true, we have barely penetrated but the most superficial of surface in this quickly emerging field. New discoveries are to be made with some high impact advances sketched above which will be complemented, like is usually in science, by unforeseen discoveries in the near future. We are very excited to be part of this collective and intriguing endeavor and agree with E. Teller when he said that
\begin{center}
    ``{\it the science of today is the technology of tomorrow..}''\;\;\;--Edward Teller.
\end{center}
With that in mind we look forward to  the future of the field of quantum materials control with the highest expectations. \\[0.5cm]

{\textbf{Acknowledgments}} 
We thank Jörg Harms for helping with some of the figures. DMK acknowledges the Deutsche Forschungsgemeinschaft (DFG, German Research Foundation) for support through RTG 1995, within the Priority Program SPP 2244 ``2DMP'' and under Germany's Excellence Strategy - Cluster of Excellence Matter and Light for Quantum Computing (ML4Q) EXC 2004/1 - 390534769.
AR acknowledges financial support from the Cluster of Excellence 'CUI: Advanced Imaging of Matter'- EXC 2056 - project ID 390715994 and  and  SFB-925 "Light induced dynamics and control of correlated quantum systems" – project 170620586  of the Deutsche Forschungsgemeinschaft (DFG), Grupos Consolidados (IT1249-19). We acknowledge support from the Max Planck-New York City Center for Non-Equilibrium Quantum Phenomena. The Flatiron Institute is a division of the Simons Foundation.

This is a preprint of the following chapter: D.M. Kennes, A. Rubio, ``A New Era of Quantum Materials Mastery and Quantum Simulators In and Out of Equilibrium''  published in Sketches of Physics: The Celebration Collection, edited by Roberta Citro, Morten Hjorth-Jensen, Maciej Lewenstein, Angel Rubio, Wolfgang P. Schleich, James D. Wells and Gary P. Zank, 2022, Springer reproduced with permission of Springer. The final authenticated version is available online at: http://dx.doi.org/[DOI shall be updated once given]

\bibliography{references_DK,references_MC,references_MS,references_JM,references_ADLT,references_SG,references_ADLT_ii,refs,refs2,ref_Natphys}

\end{document}